\documentclass[onecolumn]{aastex7}
\usepackage{amsmath,amssymb}
\usepackage{graphicx}
\usepackage{algorithm}
\usepackage{algpseudocode}
\usepackage{multirow}

\usepackage{subcaption}
\shorttitle{Fast QU fitting}

\begin{document}
\title{VROOM-SBI: A Fast Simulation-Based Bayesian Inference Methodology for Stokes QU-Fitting in Radio Interferometry}
\author{Arpan Pal}
\affiliation{National Radio Astronomy Observatory, P.O. Box O, Socorro, NM 87801, USA}
\affiliation{National Centre for Radio Astrophysics, Tata Institute of Fundamental Research, Pune, India}
\affiliation{Fulbright-Nehru Fellowship, United States-India Educational Foundation, Fulbright Commission in India}
\email{arpan522000@gmail.com}

\author{Preshanth Jagannathan}
\affiliation{National Radio Astronomy Observatory, P.O. Box O, Socorro, NM 87801, USA}
\email{pjaganna@nrao.edu}

\begin{abstract}
Bayesian QU-fitting is among the most accurate approaches for line-of-sight Faraday inference, but its per-pixel computational cost has made survey-scale application infeasible. QU-fitting is an alternative to Faraday synthesis with comparable accuracy in recovering line-of-sight Faraday components, but it has historically been computationally prohibitive at survey scale. Fitting to the Stokes spectra in $Q$ and $U$ through Bayesian inference is effective but slow. We introduce \texttt{VROOM-SBI}, which uses simulation-based inference, particularly neural posterior estimation, to speed up inference. Our results are comparable to both Faraday synthesis and QU-fitting, and deliver a speedup of $\sim$$500$ over classical QU-fitting implementations. We provide an open code repository and tools along with trained models via HuggingFace for the four standard depolarization models in common use, trained on VLA L-band frequency coverage.
\end{abstract}

\section{Introduction}

Magnetic fields alter the polarization state of radio emission either at the source or in the intervening regions between the source and the observer. The state of the polarized light at emission is encoded in the electric vector position angle (EVPA, also known as the polarization position angle). Magnetic fields and plasma along the line of sight rotate the EVPA in proportion to the strength of the magnetic field along the line of sight. This is called Faraday rotation and is a commonly observed phenomenon in radio astronomy and radio interferometry \citep{Burn1966}, \citep{BrentjensDeBruyn2005}
The polarized state of light \citep{Chandrasekhar} can be parameterized into four Stokes parameters: $I$ total intensity, $Q$ and $U$ linearly polarized intensity, and $V$ the circularly polarized intensity of the incident light. Together they provide a complete description of the polarization state of the electromagnetic radiation on the Poincar\'{e} sphere. Of particular interest is the complex linear polarization given by
\begin{equation}
P = Q + iU = pI e^{2i\chi},
\end{equation}
where $Q$ and $U$ are the Stokes parameters, $P = pI$ is the linearly polarized intensity, $p$ is the fractional linear polarization, and $\chi = \frac{1}{2}\arctan(U/Q)$ is the EVPA. The Faraday rotation induced by an intervening plasma can be written as
\begin{equation}
\phi = 0.81 \int_{\text{source}}^{\text{observer}} n_e \, B_\parallel \, dl \quad
\text{rad m}^{-2},
\label{eq:faraday_depth}
\end{equation}
where $n_e$ is the electron number density (cm$^{-3}$), $B_{\parallel}$ is the component of the magnetic field along the line of sight ($\mu$G), and $dl$ is the path length through the magnetized plasma (pc). This rotation can occur any time there is intervening plasma and is a cumulative measure that manifests as a linear relationship between the observed EVPA $\chi(\lambda^2)$ and the square of the wavelength, given by
\begin{equation}
\chi(\lambda^2) = \chi_0 + \phi\lambda^2.
\end{equation}

This relationship has been a widely used probe of the intervening magnetic field. The most successful demonstration of it was by \cite{TaylorR2009}, who used the linear polarization data of the NVSS survey \citep{NVSS1998} to derive an estimate of the rotation measure of our galaxy as a function of the line of sight toward background polarized point sources.

\cite{BrentjensDeBruyn2005} introduced Faraday synthesis, which uses the Fourier transform relationship between the linearly polarized light and the Faraday dispersion function $F(\phi)$, given by

\begin{equation}
\mathcal{P}(\lambda^2) = \int_{-\infty}^{+\infty} F(\phi) \, e^{2i\phi\lambda^2} \, d\phi.
\label{eq:faraday_transform}
\end{equation}

Inverting this equation gives the Faraday depth spectrum, an effective means of decomposing the line of sight into a series of intervening clouds of plasma. This Fourier relationship allows for CLEAN-like algorithms \citep{Heald2009} for the determination of the Faraday depth spectrum. There are many implementations, such as \texttt{pyrmsynth}, which use this approach for image cubes. \texttt{RMTools} \citep{RMTools} is a collection of tools for deriving the Faraday depth curated into a collection.

An alternative approach presented by \cite{farns2011} and \cite{OSullivan2012} is model fitting to the Stokes $Q$ and $U$ spectra of an observation. The models for the spectra were derived from detailed depolarization models from \cite{Burn1966} and \cite{Sokoloff1998}. For a more detailed overview of the models we refer the reader to Section~\ref{sec:architecture}. As both methods matured, a community-wide benchmark \citep{Sun2015} found that QU-fitting outperforms CLEAN-based Faraday synthesis in accuracy but at considerably greater computational cost. Subsequent work characterized this trade-off further, refining the treatment of non-physical structure at negative $\lambda^2$ \citep{Schnitzeler2018, Pratley2021}, benchmarking QU-fitting performance for closely spaced Faraday-complex sources \citep{Miyashita2019MNRAS, Miyashita2019Galaxies}, extending the technique to AGN magneto-ionic complexity \citep{Pasetto2021}, and validating it across additional frequency coverage relevant to next-generation surveys \citep{Oberhelman2026}.

\subsection{Computational Bottleneck}

Model fitting was inherently limited by the compute needed to fit complex models to the data, and this remains a bottleneck even as implemented in \texttt{RMTools}. For QU-fitting and error determination it takes on the order of a few minutes per spectrum along a single line of sight. A typical image from an SKA precursor instrument (VLA, uGMRT, MeerKAT) contains $4096 \times 4096$ spatial pixels, and even for a subset of active pixels above a noise threshold (see Fig.~\ref{fig:rm_comparison}), the compute time is prohibitive at roughly 1000 core-hours on a 16-core workstation, and would require a large cluster to complete in reasonable wall-clock time.

With the growing computational feasibility of ML approaches, \citet{SupervisedRM} proposed performing Faraday synthesis using neural networks trained to invert the Fourier transform in Equation~\ref{eq:faraday_transform}. That approach is innovative but targets the Fourier domain rather than the model parameter domain: training a network on a fixed frequency grid and spatial resolution is limiting, as these are the parameters most often altered during scientific analysis. Such a change requires retraining the network, upon which the amortization of the upfront compute cost is lost.

By contrast, our approach models the mapping between physically motivated model parameters and their posterior distributions. Since those parameters encode the physics of polarization and depolarization, the trained network is expected to transfer naturally to observations with frequency coverage and noise statistics similar to those used in training; verifying generalization to substantially different observing configurations is left to future work.

\section{Simulation-Based Inference}
\label{sec:sbi}

Simulation-based inference (SBI) is a class of algorithms that model the likelihood function or learn a direct mapping from model parameters to posterior distributions \citep{Cranmer2020}. In the QU-fitting context this is a natural fit: our forward simulator is cheap to evaluate. SBI lets us spend that cheap forward compute up front to cover the parameter space of interest, and then use neural posterior estimation (NPE) to learn the posterior as a conditional density estimator.

NPE trains a conditional density estimator to directly approximate the posterior $p(\theta | x)$ \citep{Papamakarios2016, Lueckmann2017, Greenberg2019}. Here, $\theta \in \mathbb{R}^d$ denotes the unknown model parameters we wish to infer, and $x \in \mathbb{R}^m$ denotes the observed (or simulated) data, where $p(\theta)$ encodes the prior information about the parameters. The simulator is a stochastic forward model $x = f(\theta) + \epsilon$, where $f(\theta)$ maps parameters to noiseless model output and $\epsilon$ represents observation noise. Given $N$ parameter samples $\{\theta_i\}_{i=1}^N$ drawn from the prior and their corresponding simulated observations $\{x_i\}_{i=1}^N$, a neural density estimator $q_{\boldsymbol{w}}(\theta | x)$, parameterized by neural network weights $\boldsymbol{w}$, is trained to approximate the true posterior $p(\theta | x)$ by minimizing the negative log-likelihood,
\begin{equation}
    \mathcal{L}(\boldsymbol{w}) = -\frac{1}{N} \sum_{i=1}^{N}
    \log q_{\boldsymbol{w}}(\theta_i | x_i),
    \label{eq:npe_loss}
\end{equation}
so that for any new observation $x_\text{obs}$ we can immediately sample from $q_{\boldsymbol{w}}(\theta | x_\text{obs})$ to obtain posterior samples without any additional likelihood evaluations. Once trained, the neural network performs inference on any new observation in milliseconds, making it ideal for survey-scale applications where millions of sources require parameter estimation.

\section{VROOM-SBI Architecture and Training}
\label{sec:architecture}

We train separate posterior estimators for known depolarization models, commonly used in RM studies, on L-band (1--2~GHz) frequency coverage spanning the typical VLA L-band configuration, covering Faraday depths $\phi \in [-500, 500]$~rad~m$^{-2}$ (Table~\ref{tab:models}).

\subsection{Physical Models and Parameters}
The functional forms and parameter bounds are summarized in Table~\ref{tab:models}

\begin{table}[ht!]
\centering
\caption{Depolarization models, functional forms, and prior bounds. Here $P(\lambda^2)$ is the complex polarized intensity as a function of wavelength-squared, $p_0$ is the intrinsic fractional polarization, $\chi_0$ is the intrinsic polarization angle, $\phi$ is the Faraday depth, $\phi_c$ is the central Faraday depth of the slab model, $\Delta\phi$ is the Faraday thickness, and $\sigma_\phi$ is the Faraday dispersion parameter.}
\begin{tabular}{lll}
\hline
Model & Functional Form & Parameters \& Bounds \\
\hline

\multirow{3}{*}{Faraday-thin}
& $P(\lambda^2) = p_0 e^{2i(\chi_0 + \phi\lambda^2)}$
& $\phi \in \left(-500, 500\right)\ \mathrm{rad\ m^{-2}}$ \\
& & $p_0 \in \left(0.01, 1.0\right)$ \\
& & $\chi_0 \in \left(0, \pi\right)$ \\

\hline

\multirow{3}{*}{Burn slab \citep{Burn1966}}
& $P(\lambda^2) = p_0 \mathrm{sinc}(\Delta\phi\lambda^2)$
& $\phi_c \in \left(-500, 500\right)\ \mathrm{rad\ m^{-2}}$ \\
& $\times e^{2i(\chi_0 + \phi_c\lambda^2)}$
& $\Delta\phi \in \left(0, 200\right)\ \mathrm{rad\ m^{-2}}$ \\
& & $p_0 \in \left(0.01, 1.0\right),\ \chi_0 \in \left(0, \pi\right)$ \\

\hline

\multirow{3}{*}{External dispersion \citep{Burn1966}}
& $P(\lambda^2) = p_0 e^{-2\sigma_\phi^2\lambda^4}$
& $\phi \in \left(-500, 500\right)\ \mathrm{rad\ m^{-2}}$ \\
& $\times e^{2i(\chi_0 + \phi\lambda^2)}$
& $\sigma_\phi \in \left(0, 200\right)\ \mathrm{rad\ m^{-2}}$ \\
& & $p_0 \in \left(0.01, 1.0\right),\ \chi_0 \in \left(0, \pi\right)$ \\

\hline

\multirow{3}{*}{Internal dispersion \citep{Sokoloff1998}}
& $P(\lambda^2) = p_0 \frac{1 - e^{-S}}{S}$
& $\phi \in \left(-500, 500\right)\ \mathrm{rad\ m^{-2}}$ \\
& where $S = 2\sigma_\phi^2\lambda^4 - 2i\phi\lambda^2$
& $\sigma_\phi \in \left(0, 200\right)\ \mathrm{rad\ m^{-2}}$ \\
& & $p_0 \in \left(0.01, 1.0\right),\ \chi_0 \in \left(0, \pi\right)$ \\

\hline
\end{tabular}
\label{tab:models}
\end{table}

For multi-component models ($N \geq 2$), the total complex polarization is constructed as the sum of the individual component spectra,
$P_{\mathrm{total}}(\lambda^2) = \sum_{j=1}^N P_j(\lambda^2)$,
where $P_j(\lambda^2)$ denotes the polarized emission from the $j$th Faraday component evaluated according to the chosen depolarization model. To remove the label-switching degeneracy inherent to exchangeable multi-component parameterizations, we impose the ordering $\phi_1 > \phi_2 > \cdots > \phi_N$, enforced through sorting during both prior sampling and inference. The number of training simulations scales empirically as $N_{\mathrm{sim}} \propto N^{2}$ with the number of components in order to maintain approximately constant posterior-volume coverage as the dimensionality of parameter space increases, with a baseline of $N_{\mathrm{base}} = 4{,}096{,}000$ simulations for the single-component case ($N = 1$). For the scope of this paper, we will discuss the performance of the trained posteriors for the the mentioned four 1-component models and also include a 2-component Faraday thin model.

\subsection{Noise Model}

The noise model is additive per channel. For each simulation we first generate per-channel inverse-variance weights by sampling a noise ratio $R = \sigma_{\mathrm{max}}/\sigma_{\mathrm{min}}$ from a log-uniform distribution (typically $R \in [2, 300]$ to cover realistic bandpass variation), then drawing per-channel noise levels $\sigma_k$ from $\log\mathcal{U}(1, R)$ (relative units). We compute the per-observation noise median $\sigma_{\mathrm{med}} = \mathrm{median}(\sigma_k)$ and define normalized inverse-variance channel weights as

\begin{equation}
w_k = \min\!\left(1,\left(\frac{\sigma_{\mathrm{med}}}{\sigma_k}\right)^2\right).
\end{equation}

Normalising by $\sigma_{\mathrm{med}}$ anchors the weight scale to the typical channel quality, so that the network always receives weights in $[0,1]$ regardless of the absolute noise level of a given observation; the $\min(1,\cdot)$ cap prevents any single anomalously clean channel from dominating the representation.

A baseline noise level $\sigma_{\mathrm{base}}$ is drawn uniformly from $[0.001, 0.05]$ per observation, and per-channel additive noise is then set to $\sigma_k^{\mathrm{obs}} = \sigma_{\mathrm{base}} / \sqrt{w_k}$ for good channels (where $w_k > 0$) and zero for flagged channels. This is physically motivated as additive radiometer-like noise, which differs from a percentage of signal based noise and ensures robustness to observations of varying absolute noise floors.

Prior samples are drawn using scrambled Sobol \citep{Sobol1967} quasi-random sequences rather than plain Monte Carlo, which we found gave noticeably more uniform coverage of the prior hypercube at fixed simulation budget.

We augment the training data with three independently applied RFI patterns: scattered missing channels (applied with probability $0.30$), contiguous gaps of $2$--$8$ channels (probability $0.30$), and large RFI blocks of $10$--$30$ contiguous channels (probability $0.10$), leaving approximately $30\%$ of training spectra with no RFI excision.

\subsection{Neural Architecture}
\label{arch}
For the QU models, the neural architecture consists of a fully connected embedding network followed by a normalizing flow. The embedding network has hidden dimensions $[256, 128]$, output dimension $64$, LayerNorm, and dropout $0.1$. This embedding is passed to a Neural Spline Flow \citep{Durkan2019} with $15$ coupling transforms, $256$ hidden features, and $16$ spline bins.

Training uses Adam optimizer at learning rate $5\times10^{-4}$ with early stopping when validation loss plateaus. At inference time, the trained flow draws $5{,}000$ posterior samples in a single batched forward pass, completing in under $0.1$~seconds on GPU.

The network input is a three-channel observation: normalized Stokes $Q$ and $U$ divided by Stokes $I$ (fractional polarization), and per-channel inverse-variance weights derived from noise estimation.

\begin{table*}[ht!]
\centering
\caption{Parameter recovery for single-component models (200 test samples). ``Cov'' is the empirical coverage at the nominal 68\% credible level.}
\label{tab:recovery_n1}
\begin{tabular}{llccc}
\hline
Model & Parameter & MedAE & Bias & Cov (\%) \\
\hline
\multirow{3}{*}{Faraday thin}
 & $\phi\ \mathrm{(rad\ m^{-2})}$        & 0.27  & $-0.07$   & 95 \\
 & $p_0$                                 & 0.013 & $+0.004$  & 62 \\
 & $\chi_0\ \mathrm{(rad)}$              & 0.014 & $+0.005$  & 93 \\
\hline
\multirow{4}{*}{Internal disp.}
 & $\phi\ \mathrm{(rad\ m^{-2})}$        & 156.0 & $-10.5$   & 69 \\
 & $\sigma_\phi\ \mathrm{(rad\ m^{-2})}$ & 24.3  & $+0.04$   & 71 \\
 & $p_0$                                 & 0.120 & $+0.007$  & 72 \\
 & $\chi_0\ \mathrm{(rad)}$              & 0.220 & $+0.023$  & 68 \\
\hline
\multirow{4}{*}{External disp.}
 & $\phi\ \mathrm{(rad\ m^{-2})}$        & 176.8 & $-1.90$   & 69 \\
 & $\sigma_\phi\ \mathrm{(rad\ m^{-2})}$ & 26.8  & $+0.23$   & 68 \\
 & $p_0$                                 & 0.233 & $+0.019$  & 69 \\
 & $\chi_0\ \mathrm{(rad)}$              & 0.682 & $-0.087$  & 69 \\
\hline
\multirow{4}{*}{Burn slab}
 & $\phi_c\ \mathrm{(rad\ m^{-2})}$      & 1.38  & $+0.21$   & 89 \\
 & $\Delta\phi\ \mathrm{(rad\ m^{-2})}$  & 1.20  & $+0.02$   & 84 \\
 & $p_0$                                 & 0.022 & $+0.002$  & 82 \\
 & $\chi_0\ \mathrm{(rad)}$              & 0.049 & $-0.001$  & 84 \\
\hline
\end{tabular}
\end{table*}
\subsection{Spectral-Shape Normalisation}
\label{sec:specnorm}

The Stokes $Q$ and $U$ spectra contain contributions from both the source total intensity spectrum and the linear polarization signal. To avoid bias in fitting from the total intensity component, we normalize $Q$ and $U$ by Stokes $I$ prior to inference. Dividing each channel by the raw $I$ spectrum propagates per-channel noise directly into $q = Q/I$ and $u = U/I$, degrading phase coherence across the band. To avoid this, we must fit the Stokes I spectrum independently and then normalize with the fitted model. We train a dedicated posterior estimator for the total-intensity spectral shape that provides a smooth, noise-marginalized model of $I(\nu)$.

The forward model is a log-log cubic polynomial (using natural logarithms throughout),
\begin{equation}
\ln \frac{F(\nu)}{F(\nu_0)} =
\alpha \ln\!\left(\frac{\nu}{\nu_0}\right) +
\beta \left[\ln\!\left(\frac{\nu}{\nu_0}\right)\right]^2 +
\gamma \left[\ln\!\left(\frac{\nu}{\nu_0}\right)\right]^3 ,
\label{eq:spectral_model}
\end{equation}
where $F(\nu)$ is the flux density at observing frequency $\nu$ and $\nu_0$ is the reference frequency (the mid-band channel), chosen so that $\ln(\nu/\nu_0)$ is centered near zero across the band. The normalization $F(\nu_0) = 1$ is enforced by construction, removing the overall flux scale as a free parameter. We adopt the convention $S_\nu \propto \nu^\alpha$ with $\alpha < 0$ for steep-spectrum synchrotron sources. The three inferred parameters are $\alpha$ (spectral index), $\beta$ (spectral curvature), and $\gamma$ (third-order curvature). Uniform priors are placed on each: $\alpha \in [-3, 1]$, $\beta \in [-1, 1]$, and $\gamma \in [-0.5, 0.5]$.

The network architecture and training procedure are identical to those of the polarimetric models. At inference time, the posterior mean spectrum $\hat{I}(\nu)$ is used as the Stokes~$I$ model:
\begin{equation}
\hat{I}(\nu) = F(\nu_0)\exp\!\left[
\alpha \ln\!\left(\frac{\nu}{\nu_0}\right) +
\beta \left(\ln\!\left(\frac{\nu}{\nu_0}\right)\right)^2 +
\gamma \left(\ln\!\left(\frac{\nu}{\nu_0}\right)\right)^3
\right].
\end{equation}
The corresponding fractional polarization spectra, $q(\nu) = Q(\nu)/\hat{I}(\nu)$ and $u(\nu) = U(\nu)/\hat{I}(\nu)$, are then used as inputs to the RM estimator.

The network backbone follows that of the polarimetric models (Section~\ref{arch}), consisting of a fully connected embedding network with hidden dimensions $[256, 128]$ and an output dimension of $64$, coupled to a Neural Spline Flow with $15$ coupling transforms. The input representation, however, differs: the spectral-shape network takes a single-channel vector of length $N_\nu$ containing the normalized total-intensity spectrum $F(\nu)/F(\nu_0)$, with flagged channels set to zero, rather than the three-channel $[q, u, w]$ input used in the polarimetric case. 

Training samples are drawn using scrambled Sobol sequences. The same RFI augmentation strategy is applied to the channel weights prior to simulation, including randomly scattered missing channels, contiguous gaps, and large RFI blocks with probabilities $0.30$, $0.30$, and $0.10$, respectively. Additive noise is sampled uniformly from $[0.001,\,0.05]$ and applied consistently across all unflagged channels.

\begin{figure*}[ht!]
\centering
\includegraphics[width=0.95\textwidth]{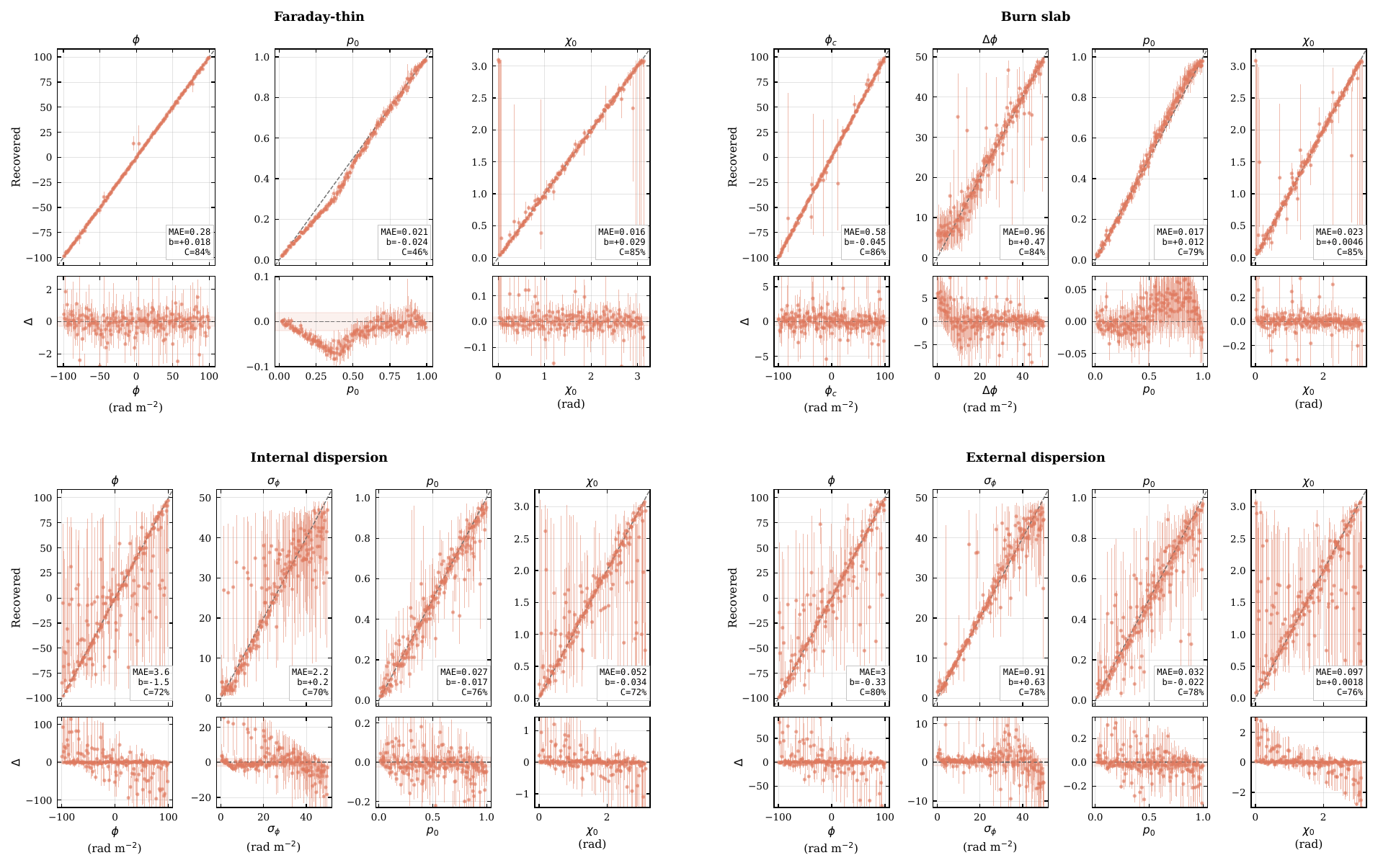}
\caption{Parameter recovery for all four single-component depolarization models (200 test cases each). \textit{Top:} posterior median versus injected truth, with 16th--84th percentile error bars; the dashed line indicates perfect recovery. \textit{Bottom:} residuals; the shaded band spans $\pm\,\mathrm{MedAE}$. Inset: MedAE, mean bias $b$, and $68\%$ credible-interval coverage $C$.}
\label{fig:recovery_all}
\end{figure*}

\section{Posterior Validation}
\label{sec:validation}

After training, we validate the trained posteriors using 200 held-out Sobol test simulations per model. For each test case, we draw $1{,}000$ posterior samples and report the posterior median as the point estimate. Table~\ref{tab:recovery_n1} summarises parameter recovery across all four single-component models. Figure~\ref{fig:recovery_all} present the full recovery results for all the models.

Coverage at the nominal $68\%$ credible level ranges from $62\%$ to $95\%$ across all parameters and models. The Faraday-thin model achieves by far the highest RM precision (MedAE $= 0.27$~rad~m$^{-2}$; coverage $95\%$), since it is the only model without a depolarization envelope competing for phase information. The internal- and external-dispersion models show substantially larger $\phi$ MedAE (156 and 177~rad~m$^{-2}$, respectively) under the wide prior, reflecting a genuine information limit rather than a training deficiency: with $\sigma_\phi$ priors extending to $200$~rad~m$^{-2}$, the dispersive depolarization envelope can erase most of the phase information tying $Q(\lambda^2)/U(\lambda^2)$ to $\phi$ well before the edge of the prior volume is reached, and the posterior correctly reports this as broad, near-nominal-coverage uncertainty ($\sim$$68$--$72\%$) rather than false precision. The fractional polarization $p_0$ in the Faraday-thin model exhibits lower coverage ($62\%$); we attribute this to the geometry of the Faraday-thin model, in which $p_0$ enters only as an overall amplitude scaling of pure sinusoids in $\lambda^2$ and provides little structural information beyond the signal level itself. In every other model, either a depolarization envelope or a multi-component beat pattern varies with $\lambda^2$, giving the network an independent handle on $p_0$; accordingly, all other models recover $p_0$ within expected ranges ($69$--$82\%$).
  
\begin{figure*}[ht!]
\centering
\includegraphics[width=0.95\textwidth]{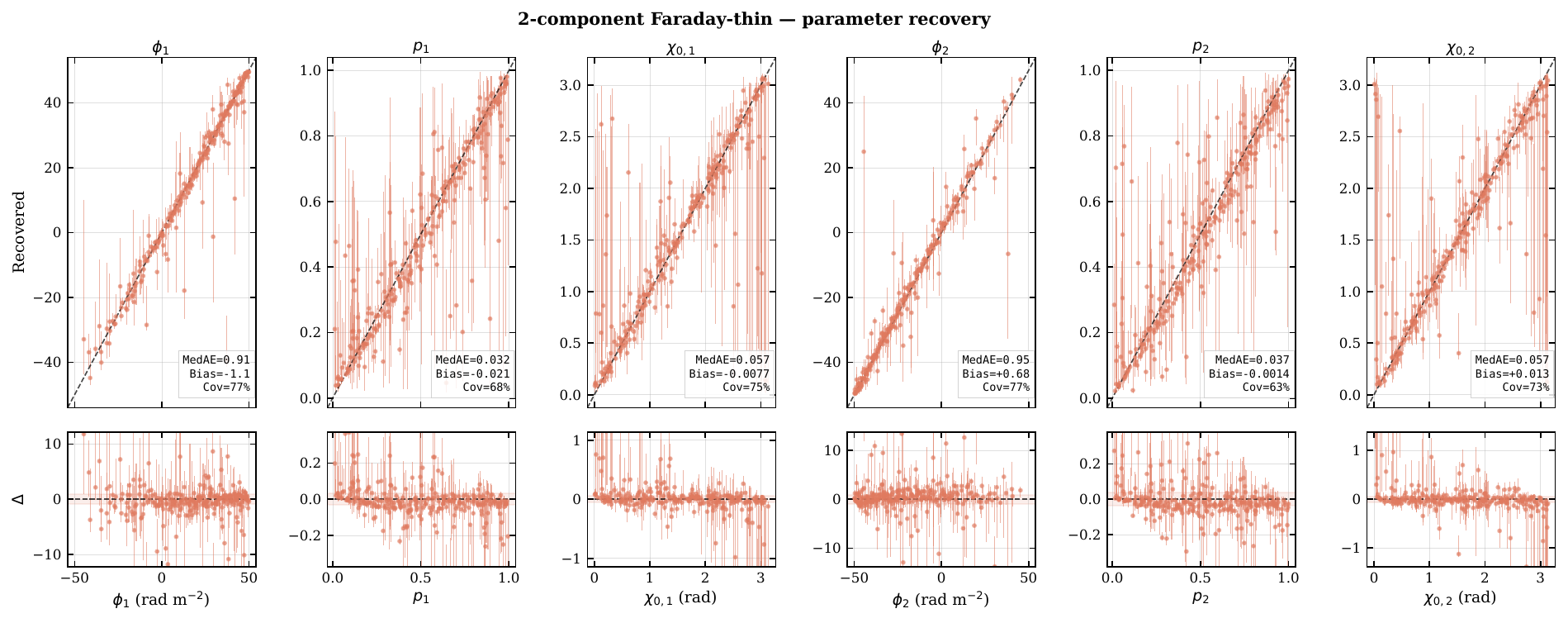}
\caption{Parameter recovery for the two-component Faraday-thin model (256 Sobol test cases). Top row: recovered (posterior median) versus true for each of the six parameters; error bars show the 16th--84th percentile interval. Bottom row: residuals. Components are ordered $\phi_1 > \phi_2$.}
\label{fig:recovery_n2}
\end{figure*}

The $\Delta\phi$--$p_0$ degeneracy in the Burn slab model reflects a limitation of the data: when the sinc null falls outside the observed $\lambda^2$ range, broader frequency coverage is required to resolve the degeneracy. Two-component recovery results are shown in Figure~\ref{fig:recovery_n2}. Both rotation measures are recovered to within $2$~rad~m$^{-2}$, with coverage close to nominal across all six parameters, confirming effective component separation. TARP joint-posterior calibration results across the full model matrix are provided in Appendix~\ref{app:sbc}.

The trained $N=2$ Faraday-thin posterior estimator recovers two-component Faraday structures from a single noisy spectrum. Figure~\ref{fig:recovery_n2} shows parameter recovery for 256 held-out Sobol test cases (6 parameters: $\phi_1$, $p_{0,1}$, $\chi_{0,1}$, $\phi_2$, $p_{0,2}$, $\chi_{0,2}$, with $\phi_1 > \phi_2$ enforced). Recovery of both rotation measures is precise (median absolute error $< 2$~rad~m$^{-2}$), with empirical coverage near the nominal 68\% level ($\phi$: 76\%, $p_0$: 65\%, $\chi_0$: 73\%). The mild under-coverage on $p_0$ is consistent with the single-component behavior.

\begin{figure*}
\centering
\includegraphics[width=\linewidth]{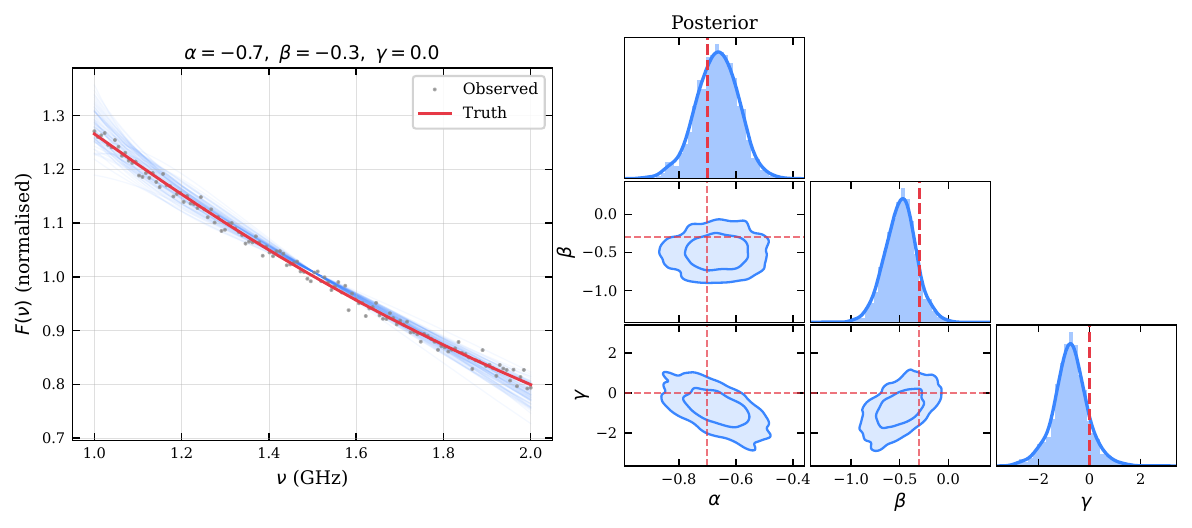}
\caption{Spectral-shape posterior for a simulated steep-spectrum source ($\alpha=-0.7$, $\beta=-0.3$, $\gamma=0.0$). \textit{Left:} Noisy observed spectrum (grey), true model (red), and 80 posterior predictive draws (blue). \textit{Right:} Marginal and joint posteriors for $\alpha$, $\beta$, and $\gamma$; contours enclose 68\% and 95\% credible regions, and red dashed lines mark the injected truth.}
\label{fig:spectral_demo}
\end{figure*}
\begin{figure*}[ht!]
\centering
\includegraphics[width=\columnwidth]{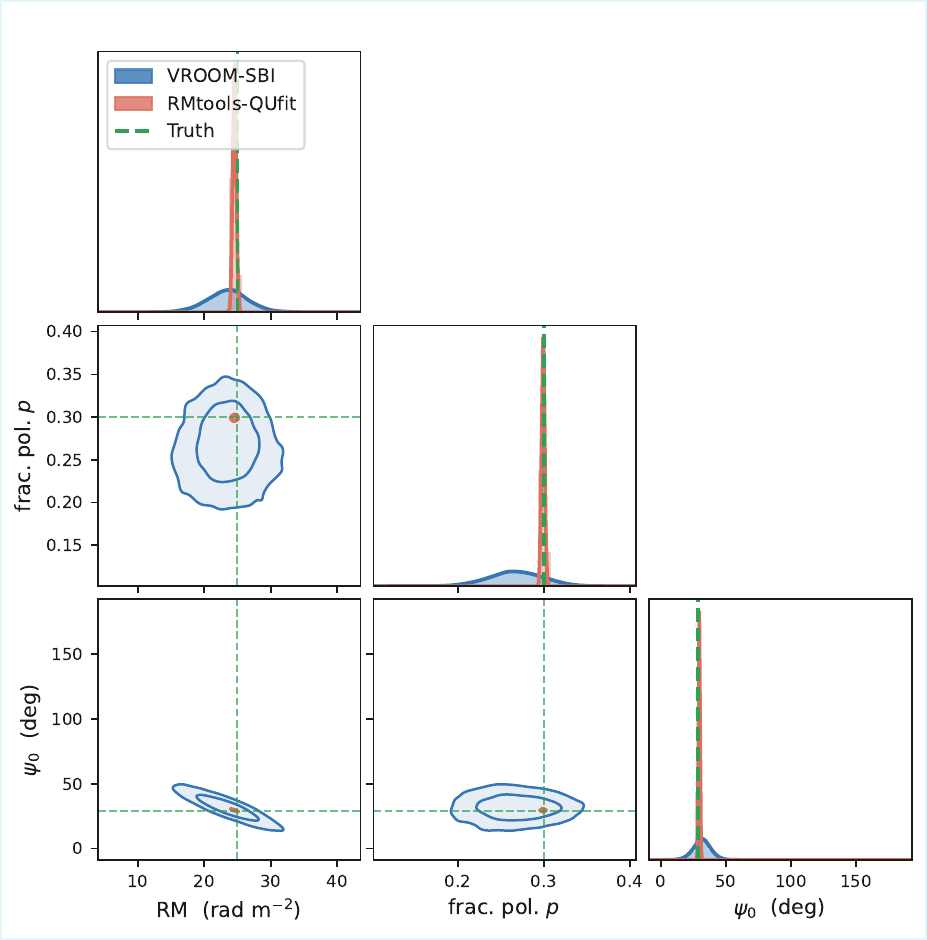}
\caption{Posterior corner plot for a simulated Faraday-thin source. Blue filled contours show the VROOM-SBI posterior (68\% and 95\% credible regions), the red curve shows the \textsc{RMtools}-QUfit posterior, and the green dashed lines mark the injected ground truth. VROOM-SBI recovers all three parameters in agreement with the truth, with somewhat broader posteriors than the likelihood-based sampler.}
\label{fig:corner_sim}
\end{figure*}
\section{Results}
\label{sec:results}

\subsection{Inference on Simulated Measurements}

We start with the spectral-shape posterior on a simulated spectrum. A source with spectral index $\alpha=-0.7$, curvature $\beta=-0.3$, and third-order curvature $\gamma=0.0$ is representative of a typical radio source. The spectrum is normalized by $F(\nu_0)$ at the reference frequency $\nu_0\approx1.5\,\mathrm{GHz}$, and independent Gaussian noise with $\sigma=0.01\,F(\nu_0)$ is added to each of the 128 channels spanning $1$--$2\,\mathrm{GHz}$, corresponding to a per-channel signal-to-noise of $\sim100$.

Figure~\ref{fig:spectral_demo} summarizes the inference. The left panel shows that the posterior predictive draws closely bracket the true model across the full band, with no systematic residual. The right panel shows the marginal and joint posteriors for $\alpha$, $\beta$, and $\gamma$. The spectral index $\alpha$ and curvature $\beta$ are both well constrained, with posteriors tightly centered on the injected values. The third-order parameter $\gamma$ is recovered with noticeably broader marginals: the $1$--$2\,\mathrm{GHz}$ octave provides limited differential leverage on a cubic log-frequency term, so the data can accommodate a range of $\gamma$ values without appreciably changing the predicted spectrum. This mild degeneracy is expected and does not affect the primary astrophysical quantities of interest; the spectral-shape model is used here to provide a smooth normalization of the polarization spectra (Section~\ref{sec:specnorm}), for which $\alpha$ and $\beta$ are the dominant terms.

Next, Figure~\ref{fig:corner_sim} shows a representative posterior corner plot for a simulated Faraday-thin source with known ground truth. Both VROOM-SBI and \textsc{RMtools}-QUfit recover all three parameters namely, rotation measure $\phi$, fractional polarization $p_0$, and intrinsic angle $\chi_0$ in agreement with the injected values. The \textsc{RMtools} posterior is narrower, as expected for a likelihood-based sampler conditioning on the exact noise model. VROOM-SBI returns a somewhat broader posterior that nevertheless encloses the truth within its 68\% credible region for all parameters. The joint $(\mathrm{RM},\,p_0)$ contour illustrates that VROOM-SBI correctly captures the mild correlation between these parameters without  any explicit likelihood evaluation.

\begin{figure*}
\centering
\includegraphics[width=0.95\textwidth]{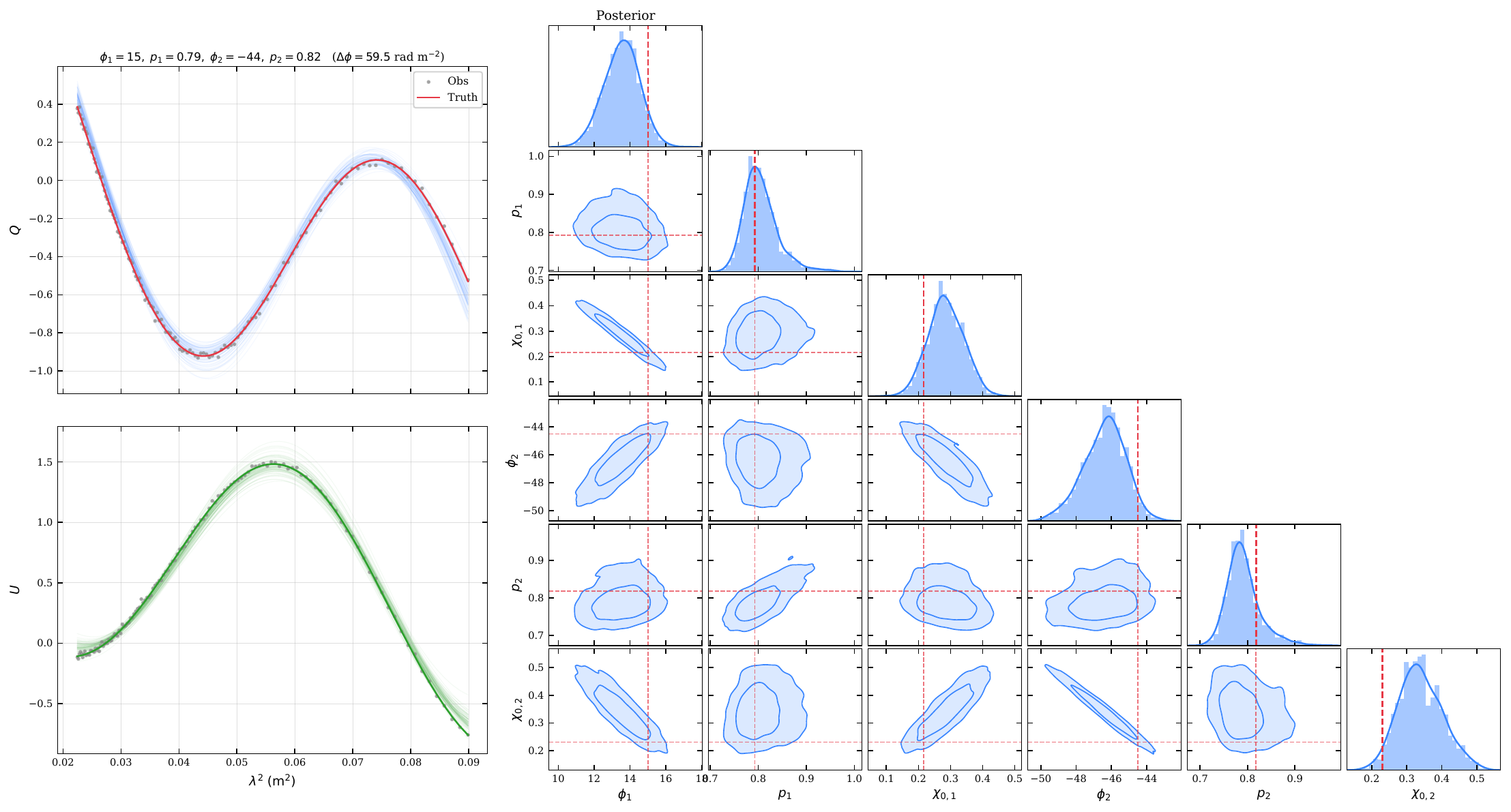}
\caption{Two-component recovery for a representative case with $\Delta\phi = 59.5$~rad~m$^{-2}$ ($\phi_1 = +15$, $\phi_2 = -44.5$~rad~m$^{-2}$). \textit{Left:} Simulated $Q$ and $U$ spectra (grey points) with posterior predictive draws (blue); the characteristic beat pattern of two interfering Faraday screens is clearly visible. \textit{Right:} Joint posterior corner plot for all six parameters; contours enclose 68\% and 95\% of the posterior mass, and red dashed lines mark the injected truth. Both rotation measures and polarization amplitudes are recovered without significant bias.}
\label{fig:n2_case}
\end{figure*}

Figure~\ref{fig:n2_case} shows a single representative two-component case with $\Delta\phi = 59.5$~rad~m$^{-2}$ between the two components. The $Q$ and $U$ spectra display the characteristic beat pattern of two interfering Faraday screens, and the network posterior correctly identifies both rotation measures, amplitudes, and intrinsic angles. The joint $(\phi_1,\phi_2)$ posterior is compact and centered on the injected truth, demonstrating that VROOM-SBI resolves closely spaced components without degeneracy when the frequency separation is sufficient.

\subsection{Inference on Real Observations}
\label{sec:real_data}

All trained posterior estimators are publicly available via \dataset[10.57967/hf/9741]{https://doi.org/10.57967/hf/9741} \citep{skunkworks_radio_astronomy_2026} on HuggingFace\footnote{\url{https://huggingface.co/skunkworks-ra/vroomsbi}}. We provide two kinds of trained posteriors, one with 128 channels and RM range of -50 to 50, while we have also provided the posteriors with 1024 channels and RM range from -500 to 500. All future releases of trained posteriors will be published in the same Huggingface repository.

VROOM-SBI provides two cube inference modes. The spectral mode (\texttt{cube-infer-spectra}) accepts a single Stokes~$I$ cube and infers the posterior over the log--log cubic spectral model (Equation~\ref{eq:spectral_model}). Active pixels are selected where the frequency-collapsed intensity $\bar{I}$ exceeds $N_\sigma \hat{\sigma}_I$, with $\hat{\sigma}_I = 1.4826 \,\mathrm{MAD}(\bar{I})$, and a default threshold of $N_\sigma = 5$.

The polarimetric mode (\texttt{cube-infer-pol}) accepts Stokes~$Q$ and $U$ cubes and, optionally, a Stokes~$I$ cube. When $I$ is provided, the spectral-shape posterior is applied to it first, yielding a smooth model $\hat{I}(\nu)$ free of per-channel noise; the fractional-polarization spectra $q = Q/\hat{I}$ and $u = U/\hat{I}$ are then formed per pixel. If no $I$ cube is supplied, $Q$ and $U$ are assumed to already be in fractional-polarization units.

Active pixels are identified in two passes. First, per-channel inverse-variance weights are computed as $w_k = \min\!\left(1, \left(\frac{\sigma_{\mathrm{med}}}{\sigma_k}\right)^2\right)$, where $\sigma_k$ is the spatial standard deviation of $Q$ in channel $k$ and $\sigma_{\mathrm{med}}$ is its median across channels. Second, the frequency-collapsed polarized intensity $\bar{P} = \left\langle \sqrt{Q_k^2 + U_k^2} \right\rangle_k$ is computed over unflagged channels, and pixels satisfying $\bar{P} \geq N_\sigma \hat{\sigma}_P$ are retained.

In both modes, the output is a set of FITS maps giving the posterior mean, standard deviation, and the 16th and 84th percentiles for each physical parameter, with unprocessed pixels set to NaN (Not-a-Number).

To demonstrate VROOM-SBI on real data, we analyze VLA L-band observations of the galaxy cluster MACS~J1752+4440, observed under project code 11B-018  (P.I.\ Annalisa Bonafede) with 3.5~hours of integration in each of the B, C, and D configurations. Calibration and imaging follow the procedure described in \citet{palgmrt}, with visibilities from all three configurations combined and imaged with \textsc{WSClean} \citep{wsclean}. The resulting spectral cubes span 1--2~GHz across 128 channels, matching the frequency setup used during training. Of these, 50 channels were excised due to RFI, leaving 78 usable channels.

\begin{figure*}[ht!]
\centering
\includegraphics[width=\textwidth]{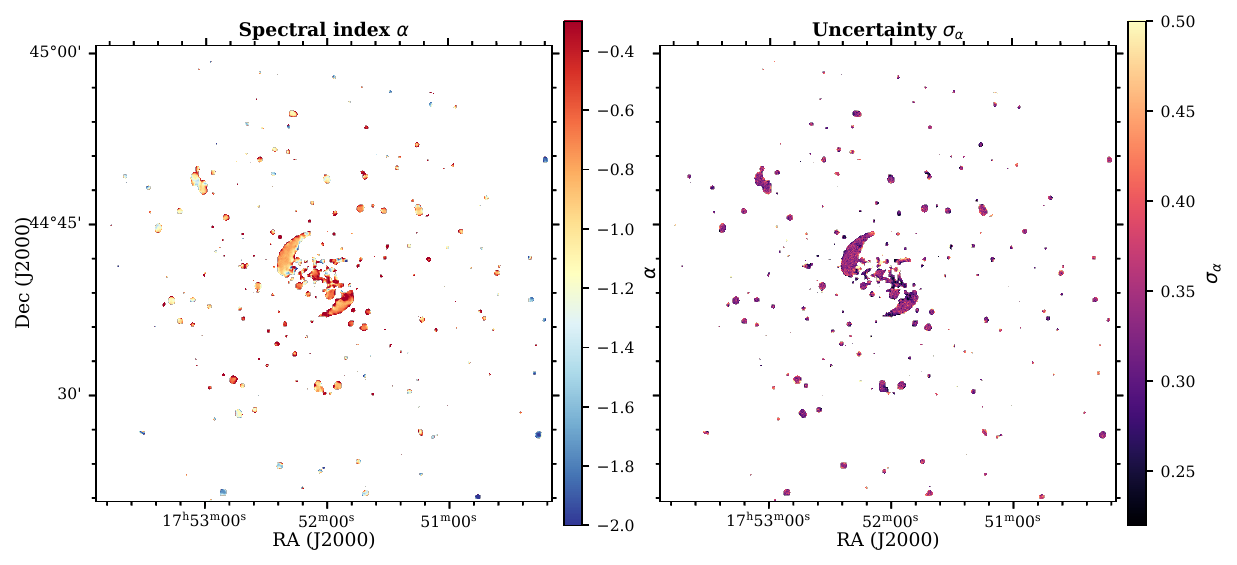}
\caption{Spectral index map of MACS~J1752+4440 derived from the VROOM-SBI total-intensity spectral-shape posterior (Section~\ref{sec:specnorm}). \textit{Left:} Posterior mean spectral index $\alpha$, clipped to $[-2.0, -0.3]$ to highlight structure in the relic. \textit{Right:} Posterior standard deviation $\sigma_\alpha$.}
\label{fig:spectral_index}
\end{figure*}

Figure~\ref{fig:spectral_index} shows the posterior mean spectral index $\alpha$ and its uncertainty $\sigma_\alpha$ across MACS~J1752+4440, derived from the VROOM-SBI total-intensity spectral-shape model applied to the same L-band cube prior to polarimetric inference. Arc-shaped radio relics are generated at merger-driven shocks through diffusive shock acceleration \citep[DSA;][]{Blandford1987}, which injects relativistic electrons with a power-law energy spectrum. As these electrons propagate downstream into the cluster, they lose energy preferentially at high frequencies through synchrotron and inverse-Compton losses, causing the radio spectrum to steepen progressively with distance from the shock front.

This hallmark flat gradient injection indices at the outermost shock edge, steepening toward the cluster center is recovered by the spectral-shape posterior. The spectral index ranges from $\alpha \approx -0.5$ at the leading edge to $\alpha \approx -1.5$ in the interior of the arc (Figure~\ref{fig:spectral_index}, left panel), consistent with the steepening profiles reported for MACS~J1752+4440 in the literature \citep{2026della}.

The uncertainty map (right panel) confirms that this gradient is statistically significant: $\sigma_\alpha < 0.30$ across the relic body, increasing only at the faint periphery where the signal-to-noise ratio is lower.

The smooth spectral-index model delivered by VROOM-SBI provides a physically motivated normalization of the polarization spectra, avoiding the per-channel noise amplification that would result from dividing by the raw noisy $I(\nu)$ spectrum directly (see Section~\ref{sec:specnorm}). VROOM-SBI identified approximately 35\,700 active pixels above a $5\sigma_P$ polarized intensity threshold and performed inference on each.

\subsection{Comparison with RM Synthesis and QU-fitting}
\label{sec:comparison}

We compare VROOM-SBI against RM synthesis, performed using the GPU-accelerated VROOM module\footnote{\url{https://github.com/skunkworks-ra/vroom}} (Pal et al., in prep.), and QU-fitting via \textsc{RMtools} \citep{RMTools} with the \textsc{dynesty} nested sampler \citep{Speagle2020} using 300 live points. Each QU-fitting pixel required approximately 2--4 minutes, making a full 35\,700-pixel run equivalent to roughly 900--1\,200 CPU-hours. Given these costs, we ran QU-fitting on a subset of 100 pixels selected by stratified sampling across the full range of polarized intensity and rotation measure.

\begin{figure*}
\centering
\includegraphics[width=\textwidth]{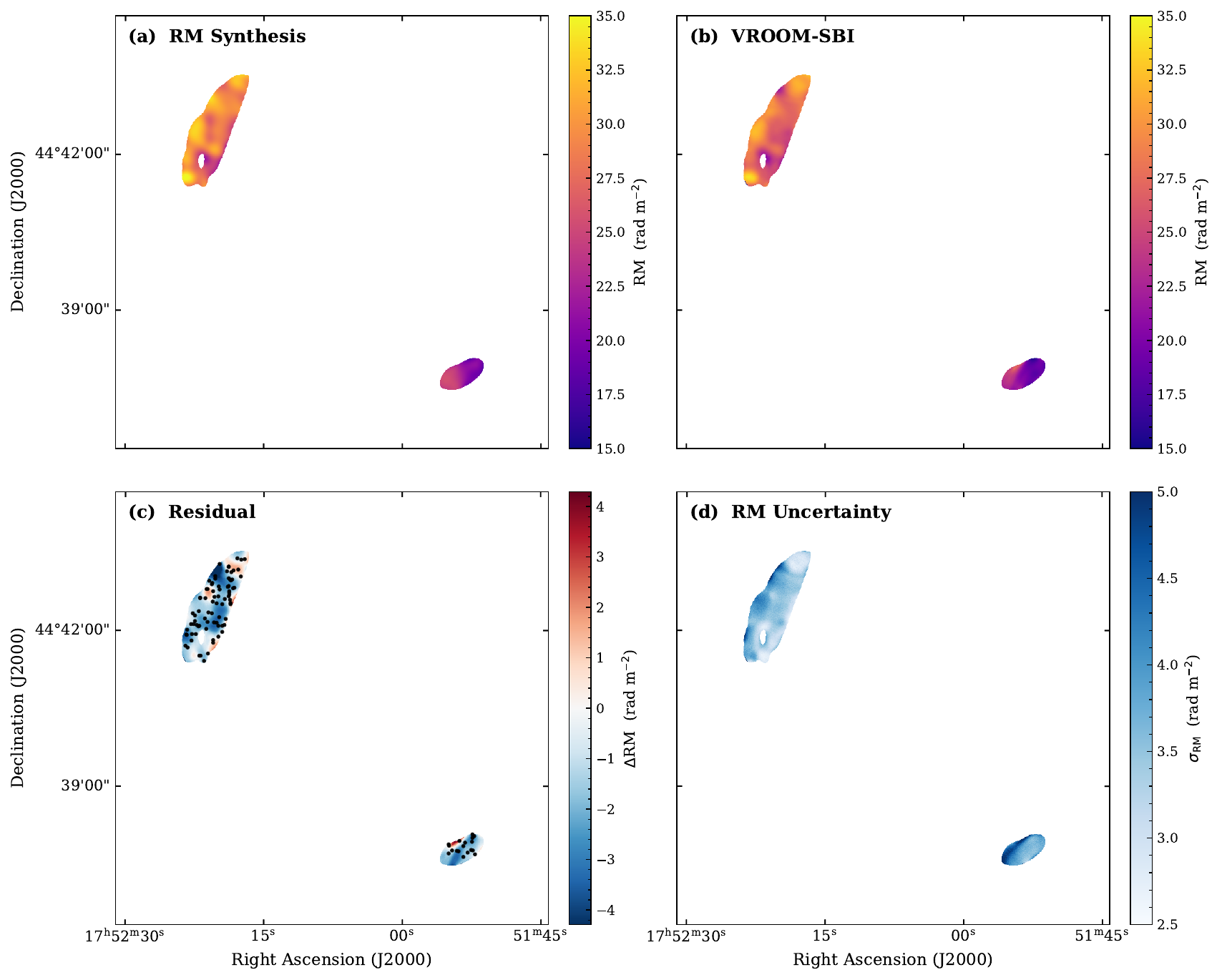}
\caption{Faraday rotation measure maps of the MACS~J1752+4440 radio relic. Panel~(a) shows the peak RM recovered by RM synthesis, and panel~(b) shows the posterior mean RM from VROOM-SBI; both share the same color scale ($15$--$35$\,rad\,m$^{-2}$). All pixels below $5\sigma_P$, where $\sigma_P$ is the noise on the polarized intensity image, are masked. Panel~(c) shows the residual between the two methods (VROOM-SBI $-$ RM synthesis); the 100 pixels selected for independent QU-fitting validation are overlaid as black points (see Section~\ref{sec:comparison}). Panel~(d) shows the posterior standard deviation of the VROOM-SBI RM, quantifying the per-pixel inference uncertainty. All maps are shown at the native angular resolution of the L-band observations.}
\label{fig:rm_comparison}
\end{figure*}

Figure~\ref{fig:rm_comparison} shows the spatial RM distribution recovered by RM synthesis and VROOM-SBI, along with their pixel-by-pixel residual. Both methods recover consistent large-scale RM structure with no systematic spatial pattern in the residual. The 100 pixels selected for QU-fitting are marked on the residual panel, confirming they sample the full extent of the source.

\begin{figure}
\centering
\includegraphics[width=\columnwidth]{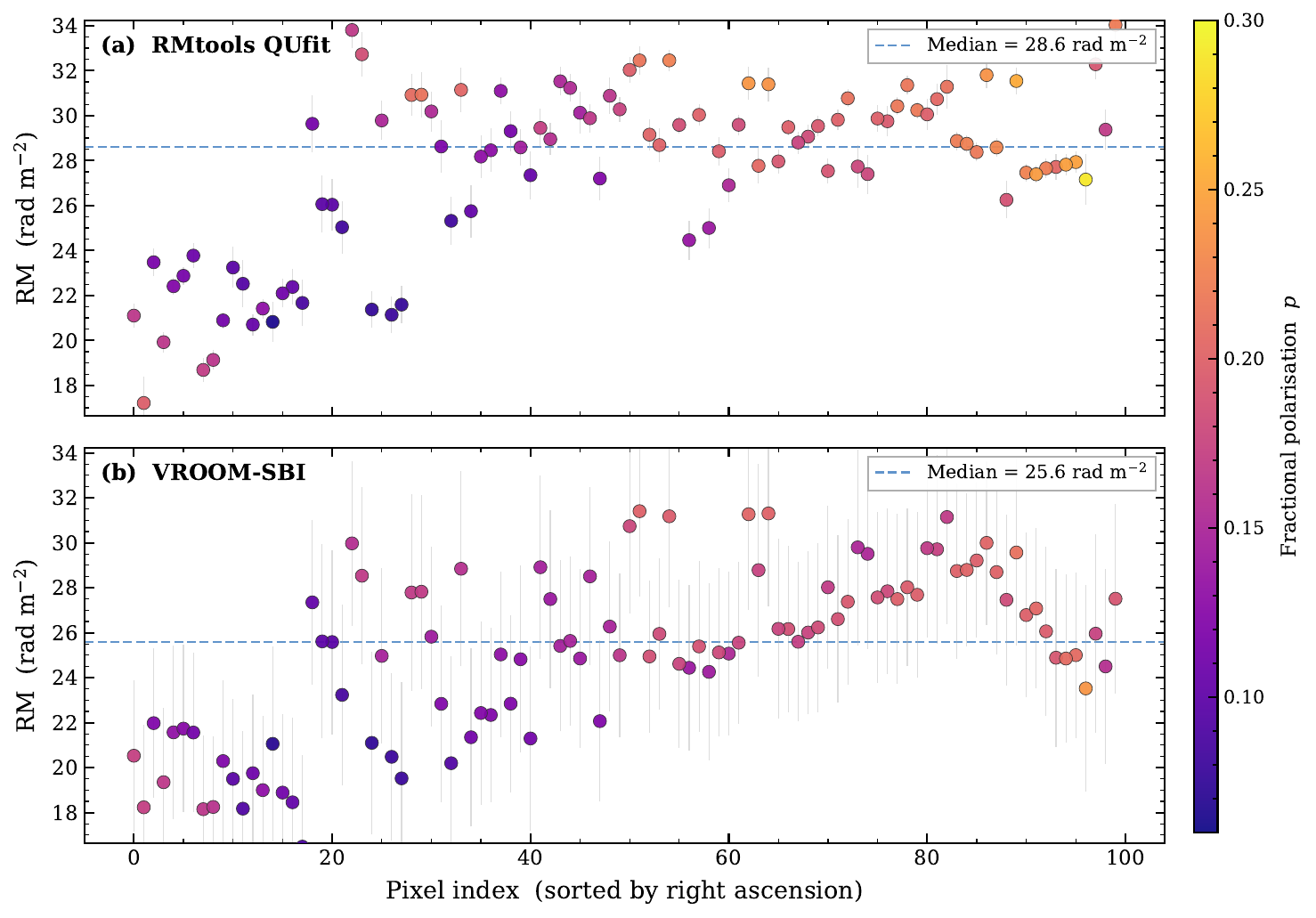}
\caption{Rotation measures recovered by \textsc{RMtools} QUfit and VROOM-SBI for 100 pixels across the MACS~J1752+4440 relic, selected by stratified sampling to span the full range of detected polarized emission. Panel~(a) shows the RM posterior mean from \textsc{RMtools} QUfit, and panel~(b) shows the corresponding VROOM-SBI result; error bars indicate the posterior standard deviation in each case. Points are colored by fractional polarization $p_0$. The dashed line in each panel marks the sample median ($28.6$ and $25.6\,\mathrm{rad\,m^{-2}}$, respectively). VROOM-SBI recovers the same large-scale RM structure as QU-fitting but with a systematic offset of $\sim2.6\,\mathrm{rad\,m^{-2}}$ and broader per-pixel posteriors.}
\label{fig:delta_rm}
\end{figure}

Figure~\ref{fig:delta_rm} shows the RM values recovered by QU-fitting and VROOM-SBI for the 100 comparison pixels. QU-fitting residuals relative to RM synthesis have a mean offset of $+0.34\,\mathrm{rad\,m^{-2}}$ (median $+0.34\,\mathrm{rad\,m^{-2}}$), consistent with the close agreement expected between a likelihood-based sampler and a direct RM synthesis estimator at high signal-to-noise. VROOM-SBI sits systematically below both RM synthesis and QU-fitting, with a mean offset of $+2.6\,\mathrm{rad\,m^{-2}}$ (median $+2.4\,\mathrm{rad\,m^{-2}}$) relative to RM synthesis and a pixel-to-pixel scatter of $1.5\,\mathrm{rad\,m^{-2}}$. The broader posterior widths of VROOM-SBI (median $\sigma_{\mathrm{RM}}=3.8\,\mathrm{rad\,m^{-2}}$ versus $0.67\,\mathrm{rad\,m^{-2}}$ for QU-fitting) reflect the amortized nature of the inference, which integrates over the full range of noise realizations seen during training rather than conditioning on the exact per-observation noise level.

\begin{figure*}[ht!]
\centering
\includegraphics[width=\columnwidth]{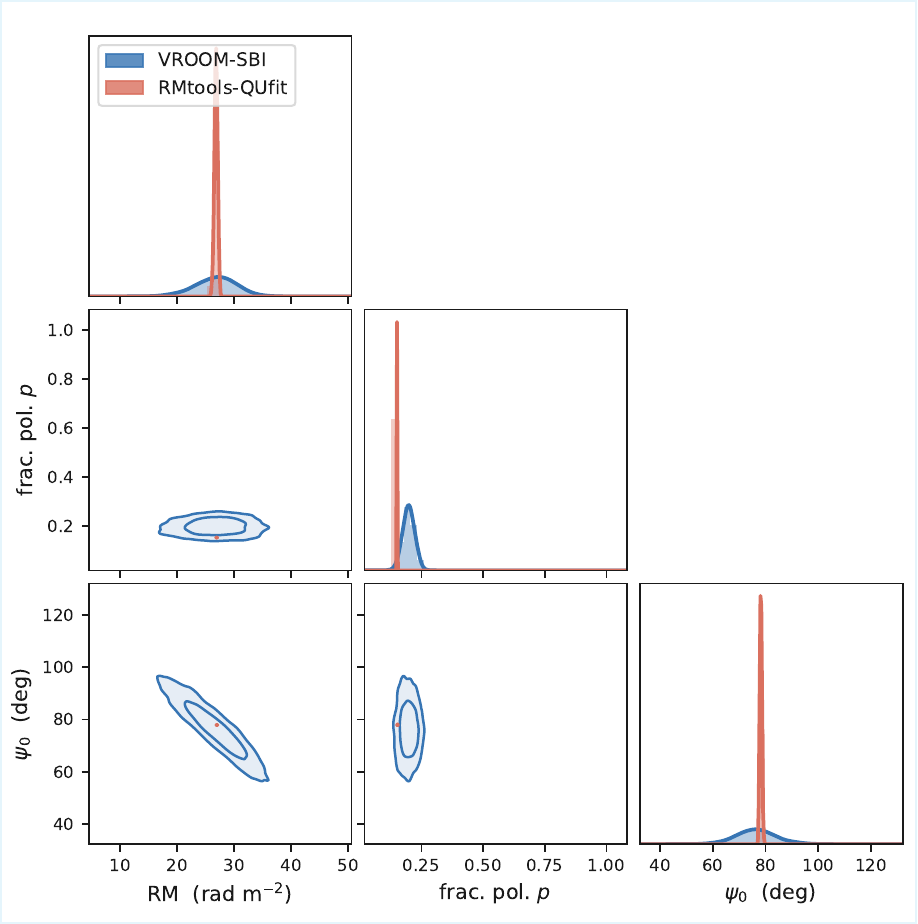}
\caption{Posterior corner plot for a representative pixel in MACS~J1752+4440 ($p_0\approx0.25$, $\mathrm{RM}\approx25\,\mathrm{rad\,m^{-2}}$). Blue contours show the VROOM-SBI posterior (68\% and 95\% credible regions); the red curve shows the \textsc{RMtools}-QUfit posterior. The \textsc{RMtools} point estimate falls within the VROOM-SBI 68\% credible region in every projection.}
\label{fig:corner_g71}
\end{figure*}

Figure~\ref{fig:corner_g71} shows the posterior comparison for a representative pixel in MACS~J1752+4440. With no ground truth available, the two methods serve as mutual validation. The rotation measure posteriors are in close agreement ($\Delta \mathrm{RM} < 1~\mathrm{rad\,m^{-2}}$), and both methods recover consistent fractional polarization and intrinsic angle $\chi_0$. The broader VROOM-SBI contours reflect the conservative nature of the amortized posterior, which must integrate over the full range of noise realizations seen during training rather than conditioning on the exact per-observation noise level as a likelihood-based sampler does. Importantly, the \textsc{RMtools} point estimate falls within the VROOM-SBI 68\% credible region in all parameter projections, indicating that the network posterior is consistent with QU-fitting for this source class.

\subsection{Computational Performance}
\label{sec:timing}

Table~\ref{tab:timing} summarizes the wall-clock cost of each method. VROOM (Pal et al. in Prep.) processes all spatial pixels without prior selection, achieving $\sim39800\,\mathrm{pixels\,s^{-1}}$ on the RTX~4060 (NUFFT + RM-CLEAN) and completing the full $4800\times4800$ field in approximately $10\,\mathrm{min}$; the $35700$ active pixels correspond to a wall-clock equivalent of under $1\,\mathrm{s}$. VROOM-SBI operates only on pixels above the $5\sigma_P$ threshold, drawing $5000$ posterior samples at $\sim0.21\,\mathrm{s\,pixel^{-1}}$ on GPU and $\sim0.59\,\mathrm{s\,pixel^{-1}}$ on CPU, giving approximately $125\,\mathrm{min}$ on GPU and $353\,\mathrm{min}$ on CPU for the full $35700$-pixel field. QU-fitting the full field would require $900$--$1200$ CPU-hours; our $100$-pixel validation subset alone took approximately $3\,\mathrm{h}$.

VROOM-SBI occupies the intermediate tier of a three-way trade-off. VROOM delivers peak RM across the full image in minutes but no parameter posteriors. QU-fitting delivers the tightest posteriors but at $\sim10^{4}$ times the per-pixel cost of VROOM-SBI. VROOM-SBI delivers posterior inference at roughly $300$ times the per-pixel cost of VROOM, making pixel-level Bayesian polarimetric analysis tractable for modern wide-field surveys where $10^{5}$--$10^{6}$ polarized sources are routinely detected.

\begin{table}
\centering
\caption{Computational cost for the MACS~J1752+4440 field ($35700$ active pixels, $78$ usable channels, $5000$ posterior samples per pixel). VROOM timing covers the full $4800\times4800$ image ($23$ million pixels); VROOM-SBI and QU-fitting cover the $35700$ active pixels only. Hardware: NVIDIA GeForce RTX~4060 (Mobile) and Intel Core Ultra~9 185H (22 cores).}
\label{tab:timing}
\begin{tabular}{lcc}
\hline
Method & Per pixel & $35700$ pixels \\
\hline
VROOM (GPU, NUFFT+RM-CLEAN) & $\sim25\,\mu\mathrm{s}$ & $\sim10\,\mathrm{min}$ ($23$M px) \\
VROOM-SBI (GPU)             & $\sim0.21\,\mathrm{s}$   & $\sim125\,\mathrm{min}$ \\
VROOM-SBI (CPU)             & $\sim0.59\,\mathrm{s}$   & $\sim353\,\mathrm{min}$ \\
QU-fitting (\textsc{dynesty}) & $1.5$--$2\,\mathrm{min}$ & $\sim900$--$1200$ CPU-hr \\
\hline
\end{tabular}
\end{table}

\subsection{Training Cost}
\label{sec:training_cost}

Table~\ref{tab:timing} covers inference only. The cost amortization has to pay back is training, and we measure it here.

Training uses streaming NPE with early stopping. One epoch of a single-component model takes $2\,\mathrm{min}\,\mathrm{32}\,\mathrm{s}$ on a single A100 GPU. The four $N=1$ posteriors stopped after $103$--$152$ epochs, so a single-component model costs $4.4$--$6.4$ hours. The full matrix of $20$ posteriors, four models at $N=1$--$5$, took $392$~GPU-hours, or $16.3$~days running continuously. Table~\ref{tab:training_cost} gives the breakdown.

Cost grows slowly with component count. Per-epoch compute goes as $N^{0.53}$ and the total cost per posterior as $N^{1.1}$, both well below the $N^{2}$ scaling of the simulation budget. The architecture is the same for every $N$ (Section~\ref{arch}), so only the parameter dimension, $3N$ for Faraday-thin and $4N$ for the rest, enters the per-epoch cost; the extra simulations are spread over more epochs rather than making each epoch more expensive.

The break-even point follows. Amortized inference is cheaper than sampling once the pixel count exceeds
\begin{equation}
N_{\mathrm{break}} = \frac{T_{\mathrm{train}}}{t_{\mathrm{QU}} - t_{\mathrm{SBI}}}
\simeq \frac{T_{\mathrm{train}}}{t_{\mathrm{QU}}},
\label{eq:breakeven}
\end{equation}
since $t_{\mathrm{SBI}} = 0.21$~s is four orders of magnitude below $t_{\mathrm{QU}} = 1.5$--$2$~min. One $N=1$ posterior pays for itself after $165$--$220$ pixels, all four after $650$--$870$, and the entire $20$-posterior matrix after $11{,}700$--$15{,}500$, a third of the $35{,}700$ pixels in the field of Section~\ref{sec:comparison}. Every field after the first is free. The comparison comes with a caveat that here GPU-hours compared against CPU-hours.

Training time is not what limits this approach as the physics gets more realistic. At $N^{1.1}$, going to $N=10$ costs a few hundred GPU-hours more and still pays for itself inside one field. What limits it is the data and the model family. A $1$--$2$~GHz band carries only so much information, which is why $\phi$ recovery for the dispersion models is poor at $N=1$ under a wide $\sigma_\phi$ prior (Table~\ref{tab:recovery_n1}); no amount of training fixes that. And with the recent advancement, the Faraday-complex structures POSSUM and LOFAR are finding are often not a sum of discrete components at all, so adding components is the wrong move regardless of budget. The answer is to change the representation, not to raise $N$: forward-model the dirty Faraday spectrum with the RMSF built in (Section~\ref{sec:conclusions}) and the dimensionality comes from the data instead of an assumed component count, while the argument above still holds.

\begin{table}
\centering
\caption{Measured training cost of the released posterior matrix, by component count $N$, averaged
over the four depolarization families. Epoch counts are recorded in the released model
checkpoints; wall-clock times are measured on a single GPU. The final row gives the total for all
$20$ posteriors.}
\label{tab:training_cost}
\begin{tabular}{ccccc}
\hline
$N$ & Epochs & Per epoch & Per posterior & Four families \\
\hline
$1$ & $103$--$152$ & $160\,\mathrm{s}$ & $5.7\,\mathrm{h}$  & $22.0\,\mathrm{h}$  \\
$2$ & $144$--$310$ & $213\,\mathrm{s}$ & $12.7\,\mathrm{h}$ & $50.8\,\mathrm{h}$  \\
$3$ & $154$--$334$ & $276\,\mathrm{s}$ & $17.8\,\mathrm{h}$ & $71.2\,\mathrm{h}$  \\
$4$ & $278$--$373$ & $334\,\mathrm{s}$ & $29.3\,\mathrm{h}$ & $117.2\,\mathrm{h}$ \\
$5$ & $148$--$481$ & $397\,\mathrm{s}$ & $32.7\,\mathrm{h}$ & $130.9\,\mathrm{h}$ \\
\hline
\multicolumn{4}{l}{Full $20$-posterior matrix} & $392\,\mathrm{h}$ \\
\hline
\end{tabular}
\end{table}

\subsection{Confidence Under Model Misspecification}
\label{sec:misspecification}

The SBC and TARP diagnostics presented above (Section~\ref{sec:validation}, Appendix~\ref{app:sbc}) establish calibration under the closed-world assumption that the true source lies within the trained physical model family and prior range. A referee raised the complementary open-world question: how does the network behave when this assumption is violated, either by genuine astrophysical Faraday complexity beyond what a given posterior was trained to represent, or by a Faraday depth outside the trained prior box?

We probe this directly using synthetic spectra generated on the same VLA L-band grid (1000--2000~MHz, 1024 channels) as the training data, using the project's own forward model (Section~\ref{sec:architecture}) rather than real data, since no publicly available Faraday-complex catalog covers this frequency range with a machine-readable spectrum: SPICE-RACS \citep{Thomson2023,Thomson2026} is confined to $\sim$800--1088~MHz, and published broadband L-band/ATCA Faraday-complex samples \citep{OSullivan2012,OSullivan2017,Pasetto2018} either use a different instrument band or report rotation measures far outside our trained prior.

\textit{Genuine complexity, within the trained prior.} We inject a two-component Faraday-thin spectrum (RM$_1=+45$, RM$_2=-140\,\mathrm{rad\,m^{-2}}$, both well inside the trained $\pm500\,\mathrm{rad\,m^{-2}}$ range) and run inference against the full matrix of 20 trained posteriors (4 physical models $\times$ $N=1$--$5$ components), selecting the best-supported model by Bayesian evidence rather than by the (currently unused) classifier. Evidence-based selection correctly identifies the injected family and component count (\texttt{faraday\_thin}, $N=2$; log evidence $3.31$, versus $1.78$ for the next-best $N=1$ fit and $\leq-0.2$ for all other families), and both injected RM values fall inside their respective 95\% posterior credible intervals (RM$_1$: $[39.9,50.9]$; RM$_2$: $[-163.3,-129.8]$). This confirms that evidence comparison across the full model matrix, not just within a fixed family, recovers genuine Faraday complexity correctly when it lies within the trained domain.

\textit{Faraday depth beyond the trained prior.} We inject a single Faraday-thin component at RM$=-900\,\mathrm{rad\,m^{-2}}$, outside the trained $\pm500\,\mathrm{rad\,m^{-2}}$ box, and run inference under the correctly-specified family (\texttt{faraday\_thin}, $N=1$) to isolate the network's behavior at the prior edge from any confound with model selection. Rather than confidently reporting a wrong point estimate, or cleanly saturating at the $-500\,\mathrm{rad\,m^{-2}}$ boundary, the posterior broadens substantially and shifts toward, but does not reach, the boundary (mean $-212$, median $-246$, $\sigma=202\,\mathrm{rad\,m^{-2}}$, 95\% CI $[-482,316]$). The posterior standard deviation is roughly $50$ times larger than typical in-prior values for this model (cf. Table~\ref{tab:recovery_n1}), which is itself an actionable signal: an anomalously broad posterior width flags the observation as poorly described by the assumed model, even without an explicit misspecification statistic.

\textit{Faraday complexity beyond the trained prior.} We repeat this test for depolarization strength rather than depth, injecting an external-dispersion component with $\sigma_\phi=400\,\mathrm{rad\,m^{-2}}$, outside the trained $[0,200]\,\mathrm{rad\,m^{-2}}$ range. The response is qualitatively similar but quantitatively milder: the posterior mean ($135.5\,\mathrm{rad\,m^{-2}}$) sits below the truth and short of the $200\,\mathrm{rad\,m^{-2}}$ edge, with $\sigma=37.2\,\mathrm{rad\,m^{-2}}$ -- broader than in-prior fits but not as extreme as the RM case, likely because strong dispersion depolarizes the spectrum toward the noise floor well before $\lambda^2_{\max}$, removing much of the leverage needed to distinguish very large $\sigma_\phi$ values from moderately large ones.

Together these tests show that VROOM-SBI does not silently extrapolate outside its trained domain with false confidence: out-of-prior truths are met with markedly inflated posterior uncertainty rather than a confidently wrong answer. This is a weaker guarantee than a dedicated misspecification statistic (e.g., a posterior-predictive check on the residual spectrum), and we do not claim the posterior width alone is a calibrated detection threshold -- establishing that would require a null distribution of widths under the trained prior, which we leave to future work. But it demonstrates the qualitative behavior a survey pipeline needs: anomalously broad or boundary-hugging posteriors are a visible, low-cost signal that a given observation may be poorly described by the assumed physical model, complementing evidence-based selection across the trained model matrix for observations that remain within it.

\subsection{Limitations and Future Work}
\label{sec:limitations}

Primary-beam leakage (polarization impurity introduced by the telescope's
primary beam response, as opposed to the point-spread function of the
interferometer) is not included in the current forward model. Adding it is
straightforward in principle, since SBI requires only that the effect be
simulable: what is needed is a prior on the beam itself, i.e.\ the telescope type, the peak leakage amplitude, and the spatial structure of  the leakage across the field of view. We are implementing this now.

A second limitation is that we do not treat Faraday complexity and spatial
extent together. Normalizing each spectrum by its own fitted $\hat{I}(\nu)$
(Section~\ref{sec:specnorm}) amounts to treating every pixel as an independent, unresolved line of sight, with no explicit model of the source's extent. The consistent alternative is to perform the Faraday fit within the deconvolution, with the source represented in a suitable spatial basis, e.g.\ Gaussian components as in multi-scale CLEAN. Nothing in the present formulation obstructs this: because inference is amortized over a model family, no per-source likelihood evaluation is required. We will pursue this in a follow-up paper.

\section{Conclusions}
\label{sec:conclusions}

We have presented \texttt{VROOM-SBI}, a simulation-based inference framework for broadband radio polarimetry that uses neural posterior estimation to deliver amortized Bayesian inference on Stokes $Q$ and $U$ spectra at survey scale. Applied to VLA L-band observations of MACS~J1752+4440, the framework recovers rotation measures broadly consistent with both RM synthesis and QU-fitting, with VROOM-SBI sitting systematically below QU-fitting by a median of $2.3\,\mathrm{rad\,m^{-2}}$ across the $100$ comparison pixels. It processes $35700$ pixels with full posterior distributions in approximately $125\,\mathrm{min}$ on a single GPU, where the equivalent QU-fitting run would require $900$--$1200$ CPU-hours---a wall-clock speedup of $\sim500\times$---while retaining per-pixel posterior uncertainty maps that RM synthesis cannot provide. The network is robust to aggressive RFI flagging, producing consistent results with $50$ of $128$ channels excised. We also present a dedicated posterior estimator for the total-intensity spectral shape which provides amortized inference of spectral parameters across the band, delivering a smooth, noise-marginalized model of $I(\nu)$ that serves as a physically motivated normalization for the polarization spectra.

These results demonstrate that full Bayesian polarimetric inference is no longer computationally prohibitive at survey scale. The SKA and its precursors (ASKAP, MeerKAT, and the ngVLA) will routinely detect $10^6$--$10^7$ polarized sources, and exploiting that data fully requires inference methods that are both statistically rigorous and computationally tractable. Polarimetric spectral fitting is the ideal domain for SBI: the forward model is analytically known and cheap to evaluate, the posterior is expensive, and the amortization is more powerful than it first appears. The trained network is amortized over every observation taken with that instrument in the same frequency configuration and with similar noise statistics as the training set. Once trained, all subsequent observations in that configuration are inferred at GPU inference cost. With \texttt{VROOM-SBI}, full Bayesian polarimetric analysis of a million-source catalog is achievable in approximately one day on a single GPU.

The radio astronomy community has historically traded accuracy for speed in Faraday analysis. RM synthesis is fast but produces a dirty Faraday spectrum convolved with the RMSF, analogous to a dirty image in aperture synthesis. Spectral model fitting is correct but slow. A natural next step, following \citet{EnsslinBell2012}, is to work in the dirty Faraday spectrum domain with a forward model that accounts for the RMSF explicitly, in the same way that \citet{RauCornwell2011} and \citet{OffringaSmirnov2017} brought spectral deconvolution into imaging through MTMFS.
If MTMFS is fast enough to be the default imaging algorithm for modern surveys, then a network that operates on the dirty Faraday spectrum and delivers correct posteriors at MTMFS speeds is fast enough for survey-scale 3D Faraday synthesis. We outline this as a direction for future work. Pre-trained models, training infrastructure, and code are publicly available; we welcome community use and contributions.

\section*{Acknowledgments}

This research was supported by the Fulbright-Nehru Doctoral Research Fellowship, funded by the United States-India Educational Foundation (USIEF) and the Fulbright Program. This work used Jetstream2 at Indiana University through allocation PHY260060 from the Advanced Cyberinfrastructure Coordination Ecosystem: Services \& Support (ACCESS) program, which is supported by U.S. National Science Foundation grants \#2138259, \#2138286, \#2138307, \#2137603, and \#2138296. We acknowledge the ACCESS program \citep{Boerner2023} and the Jetstream2 team at Indiana University for providing the computational resources and cyberinfrastructure support that enabled this work. Parts of the code development and debugging were assisted by Claude (Anthropic), used as an AI coding assistant. The National Radio Astronomy Observatory and Green Bank Observatory are facilities of the U.S. National
Science Foundation operated under cooperative agreement by Associated Universities, Inc.

\appendix
\section{Posterior Calibration: TARP}
\label{app:sbc}

We assess posterior calibration using Tests of Accuracy with Random Points \citep[TARP;][]{Lemos2023}, which verifies that credible regions of the \emph{joint} posterior have the coverage they claim: for many random reference points in parameter space, TARP compares the fraction of posterior mass closer to the reference than the true parameter against the nominal credible level, producing an empirical coverage curve (the ATRC) that should track the diagonal for a calibrated posterior. We report the signed coverage gap (positive = over-confident, negative = under-confident), the unsigned deviation area from the ideal diagonal, and a qualitative verdict for each posterior.

We initially also computed marginal rank-histogram SBC \citep{Talts2018}, the more commonly used diagnostic in the SBI literature, but do not use it as our primary calibration evidence here. SBC tests each parameter's rank uniformity independently, one histogram per parameter, and is blind to how the posterior behaves jointly across parameters. Our depolarization models have several strongly correlated or partially degenerate parameter pairs by construction (e.g.\ the $\Delta\phi$--$p_0$ degeneracy in the Burn slab model, and the amplitude--geometry coupling in Faraday-thin $p_0$ discussed in Section~\ref{sec:validation}); a posterior that correctly captures such a correlation can still show a skewed marginal rank histogram for either parameter individually, because the marginal projection of a correctly-shaped joint posterior is not itself required to be symmetric under the prior. In practice we found the SBC histograms and their KS/C2ST tests flagged the majority of parameters across the matrix as formal failures, including for posteriors whose TARP coverage curves track the diagonal closely and whose point-estimate recovery (Table~\ref{tab:recovery_n1}) is unbiased. Per-parameter SBC is also a higher-variance test at fixed sample size: because it evaluates uniformity on a 1-D histogram per parameter rather than pooling information across the full joint distribution, it is more prone to detecting statistically significant but practically small deviations, and to being distorted by the ordering constraint we impose on multi-component models ($\phi_1 > \phi_2 > \cdots$), which by construction breaks exchangeability and skews the marginal rank distribution of the ordered parameters even when the joint posterior is well calibrated. For these reasons we adopt TARP as the calibration diagnostic reported here, consistent with its design purpose of directly testing joint-posterior coverage rather than a per-parameter proxy for it \citep{Lemos2023}.

We ran TARP across the full trained model matrix: four depolarization families (Faraday-thin, Burn slab, external dispersion, internal dispersion) at each of $N=1$--$5$ components, using the currently shipped (wide-prior) posteriors listed in Table~\ref{tab:models}. Table~\ref{tab:tarp_matrix} summarizes the result for all 20 posteriors.

\begin{figure*}[ht!]
\centering
\includegraphics[width=0.31\textwidth]{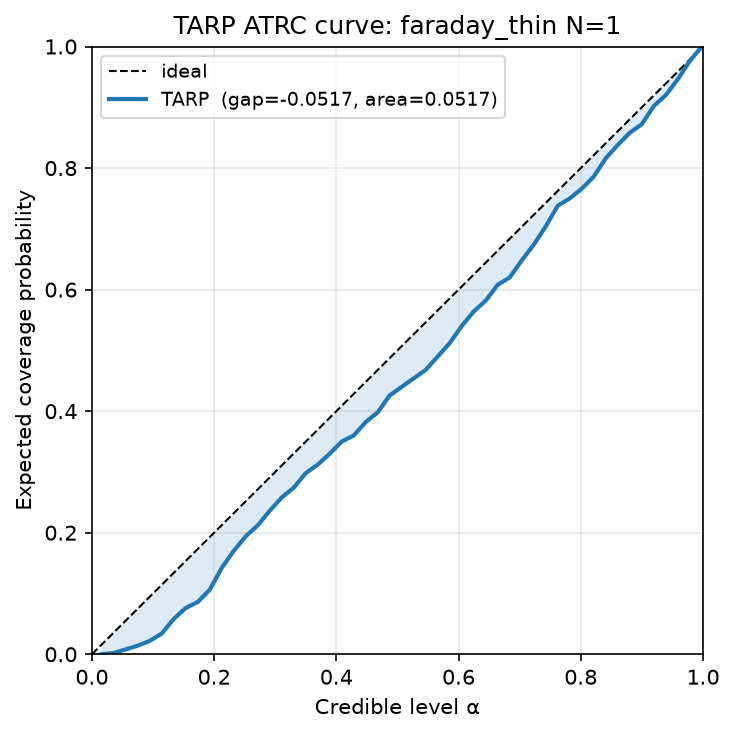}
\includegraphics[width=0.31\textwidth]{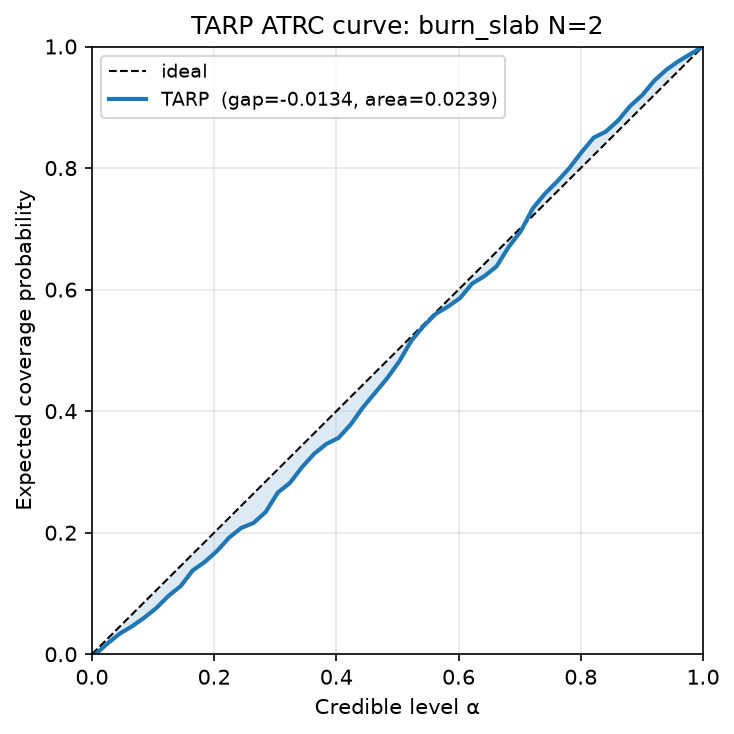}
\includegraphics[width=0.31\textwidth]{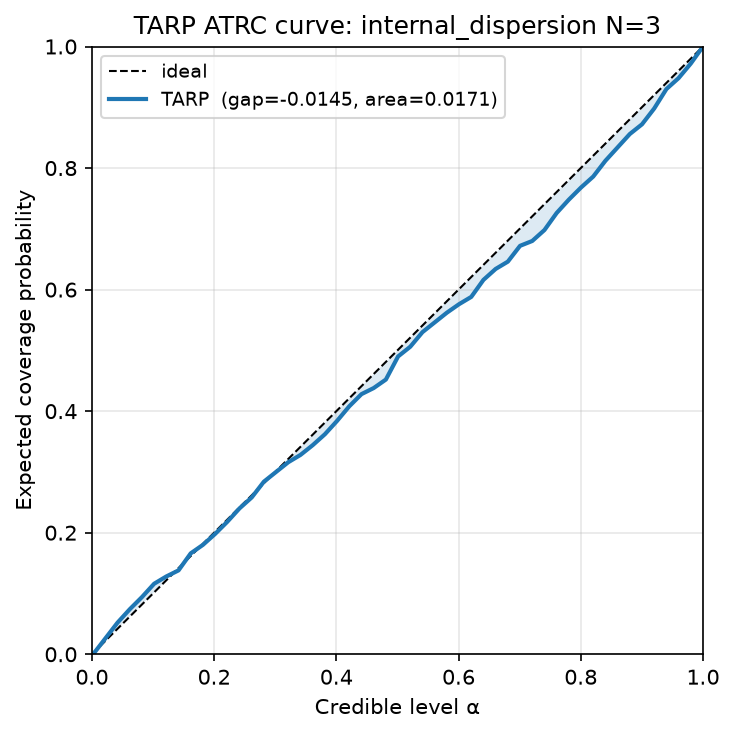}
\caption{Representative TARP ATRC curves. \textit{Left:} Faraday-thin $N=1$, the largest deviation in the matrix and formally flagged under-confident, yet within $\lesssim5\%$ coverage of the ideal diagonal at every credible level. \textit{Center:} Burn slab $N=2$, a typical calibrated example. \textit{Right:} internal dispersion $N=3$, calibrated at the information floor set by the noise model. Shaded regions show the deviation from the ideal diagonal (dashed).}
\label{fig:tarp_examples}
\end{figure*}

Fifteen of the twenty posteriors are TARP-calibrated, including every external-dispersion and internal-dispersion model at all five component counts, and the Burn slab family at $N=2$--$5$. For these, the residual coverage deviation is at or near the information floor set by the noise model itself, rather than a fixable calibration defect. The remaining five cases cluster in the Faraday-thin family ($N=1$, $2$, $4$ show a scale-type over- or under-confidence; $N=3$ shows a shape deviation) and in Burn slab $N=1$ (a shape deviation). In every case, including the flagged ones, the magnitude of the coverage deviation is small: the largest unsigned area in the matrix is $0.055$, corresponding to a deviation of a few percentage points in coverage probability at any given credible level. Figure~\ref{fig:tarp_examples} shows three representative ATRC curves spanning this range: two clean examples (Burn slab $N=2$ and internal dispersion $N=3$) and the Faraday-thin $N=1$ case, which is formally the largest deviation in the matrix but visibly tracks the ideal diagonal to within $\lesssim 5\%$ coverage across the full credible-level range.

\begin{deluxetable}{llccc}
\tablecaption{TARP joint-calibration summary across the full posterior matrix (500 held-out test cases, 500 posterior draws per case). \label{tab:tarp_matrix}}
\tablehead{
\colhead{Model} & \colhead{$N$} & \colhead{Signed gap} & \colhead{$|$Area$|$} & \colhead{Verdict}
}
\startdata
Faraday-thin & 1 & $-0.052$ & $0.052$ & under-confident\\
Faraday-thin & 2 & $+0.027$ & $0.055$ & over-confident \\
Faraday-thin & 3 & $-0.008$ & $0.030$ & mixed \\         
Faraday-thin & 4 & $-0.036$ & $0.036$ & under-confident\\
Faraday-thin & 5 & $-0.003$ & $0.011$ & calibrated \\
Burn slab & 1 & $-0.006$ & $0.027$ & mixed   \\
Burn slab & 2 & $-0.013$ & $0.024$ & calibrated \\
Burn slab & 3 & $-0.012$ & $0.016$ & calibrated  \\
Burn slab & 4 & $-0.018$ & $0.022$ & calibrated  \\
Burn slab & 5 & $-0.020$ & $0.021$ & calibrated  \\
External dispersion & 1 & $-0.001$ & $0.013$ & calibrated  \\
External dispersion & 2 & $-0.013$ & $0.016$ & calibrated  \\
External dispersion & 3 & $-0.008$ & $0.016$ & calibrated  \\
External dispersion & 4 & $-0.010$ & $0.013$ & calibrated  \\
External dispersion & 5 & $+0.001$ & $0.011$ & calibrated  \\
Internal dispersion & 1 & $-0.002$ & $0.006$ & calibrated  \\
Internal dispersion & 2 & $-0.012$ & $0.025$ & calibrated  \\
Internal dispersion & 3 & $-0.015$ & $0.017$ & calibrated  \\
Internal dispersion & 4 & $-0.014$ & $0.015$ & calibrated  \\
Internal dispersion & 5 & $-0.003$ & $0.010$ & calibrated  \\
\enddata
\end{deluxetable}

\bibliographystyle{aasjournalv7}
\bibliography{vroom}

\end{document}